\documentclass[12pt]{iopart}

\usepackage{iopams}
\usepackage{tensor}
\usepackage[dvips]{graphicx}

\renewcommand{\t}[2]{\ensuremath{\mathbf{#1}^{#2}}}	
	
\def\bra<#1|{\mathinner{\langle\,{#1}\,\vert}} 
\def\ket|#1>{\mathinner{\vert\,{#1}\,\rangle}} 
\def\red|#1|{\mathinner{\!\vert\,{#1}\,\vert\!}}
\def\braket<#1>{\mathinner{\langle\,{#1}\,\rangle}} 
\newcommand{\aanni}[1]{\ensuremath{\t{a}{#1}}}
\newcommand{\acreate}[1]{\ensuremath{\t{{a^{\dag}}}{#1}}}
\newcommand{\del}{\mathop{\delta}}
\newcommand{\half}
	{ {\frac{1}{2}}}

\newcommand{\mhalf}
	{{-\frac{1}{2}}}
\newcommand{\threej}[6]{
\ensuremath{
\Biggl(\!\!
\begin{array}{ccc}
#1 & #2 & #3 \\
#4 & #5 & #6 \\
\end{array}
\!\!\Biggr)
}}
\newcommand{\sixj}[6]{
\ensuremath{
\Biggl\{\!\!
\begin{array}{ccc}
#1 & #2 & #3 \\
#4 & #5 & #6 \\
\end{array}
\!\!\Biggr\}
}}
\newcommand{\ninej}[9]{
\ensuremath{
\left\{\!\!
\begin{array}{ccc}
#1 & #2 & #3 \\
#4 & #5 & #6 \\
#7 & #8 & #9 \\
\end{array}
\!\!\right\}
}}

\begin{document}
\title[Photodetachment intensities]{%
On the fine structure photodetachment intensities using the irreducible tensorial expression of second quantization operators
}
\author{%
O. Scharf and M.~R. Godefroid
}
\address{%
Chimie quantique et Photophysique, CP160/09, Universit\'e Libre de Bruxelles, \\
Avenue F.D. Roosevelt 50, B--1050 Bruxelles, Belgium.
}
\ead{mrgodef@ulb.ac.be,oscharf@ulb.ac.be}
%
%
\begin{abstract}
The fine-structure relative intensities of photodetachment in S$^-$ at the vicinity of the threshold have been calculated recently~\cite{Bloetal:06a} to analyze the microscope photodetachment images produced by the $s$-photoelectron. The branching ratios were obtained using the electric dipole approximation and the  standard irreducible tensorial operator techniques. The same authors observed that these relative intensities were consistent with the Cox-Engelking-Lineberger formula \cite{EngLin:79a} derived from the fractional parentage approach~\cite{Cox:75a}, in which the laser photon annihilates one of the $p$-electrons of the negative ion to promote it into the $s$-continuum. This agreement between the two formalisms was qualified as {\it remarkable}.

With this paper, we show that this agreement is understood from a {\it general} interesting angular momentum  expression relating a weighted sum of squared $9j$-symbols and a weighted sum of products of squared $6j$-symbols. 
We point out that the ``standard'' approach result is  a special case of Pan and Starace's parametrization~\cite{PanSta:93a} of the photodetachment cross sections in the term-independent approximation. The link with the Cox-Engelking-Lineberger result established in their work  makes the agreement between the standard and the fractional parentage methods even more natural. 
The present work provides another elegant and deep link between the two formalisms thanks to the irreducible tensorial expression of the second quantization form of the electric dipole transition operator. Indeed, the $(SL)J$-coupled form of the latter reproduces Pan and Starace's cross section expression from which the standard result can be derived, while the $9j$-coefficient characterizing the fractional parentage Cox-Engelking-Lineberger formula quickly emerges  when using its $(jj)J$ coupling form.
\par
\end{abstract}
\vspace*{1cm}

\maketitle
%
%

\section{Introduction}

The fine structure of S and S$^-$ have been measured recently using the photodetachment microscope 
technique \cite{Bloetal:06a}. In the Appendix of that work, branching ratios of the fine structure 
components for s-wave photodetachment of $^{32}$S$^-$ were calculated using a ``standard'' approach 
and compared with the results deduced from the ``fractional parentage'' formula of Engelking 
and Lineberger~\cite{EngLin:79a} based on Cox's treatment \cite{Coxetal:72a,Cox:75a}. As resumed in section~2, 
the two formalisms yield the same numerical results. A similar agreement between the two approaches, 
that appears as a ``surprise'' at first sight of the rather different expressions, 
was found in the study of the relative intensities of the hyperfine components of 
photodetachment from $^{17}$O$^-$  \cite{Bloetal:01a}. We show how the two formulae can 
be related to each other through an interesting general angular momentum algebra relation 
that is proven in the Appendix using a graphical approach. In their analysis, the authors of
both publications \cite{Bloetal:01a,Bloetal:06a} made no mention of the important work of 
Pan and Starace \cite{PanSta:93a}. Yet, the latter does integrate the ``standard'' approach 
formula in the case of photodetachment of a $p$-subshell electron for which limiting the partial waves 
summation to the lowest value ($l=0$) according to Wigner's threshold 
law~\cite{Wig:48a}, leads to a complete separation of dynamical and 
geometric factors. Moreover, a link between their general expression of the partial photoionization 
cross section and the previously published results \cite{Cox:75a,EngLin:79a,Schetal:80a,BerGoo:79a,GooBer:91a} 
was already established in~\cite{PanSta:93a}.  Since  Pan and Starace's contribution 
escaped to the attention of authors of recent publications on photodetachment and since the mention 
of the link with the Cox-Engelking-Lineberger results was limited in~\cite{PanSta:93a} to a rather short statement 
accompanied by a brief footnote, it is worthwhile to investigate 
Blondel {\it et al}'s ``surprise'' adopting Pan and Starace's point of view. 
This is done in section~3. In section~4, we first show how the Pan and Starace's 
cross section expression  can be derived adopting the irreducible tensorial expression 
of the second quantization form of the electric dipole transition operator. We then show 
that the $9j$-coefficient, characterizing the fractional parentage Cox-Engelking-Lineberger 
formula, emerges naturally from the recoupling of the annihilation and creation operators, 
from $(SL)J$ to $(jj)J$ coupling. 
\vspace*{1cm}

\section{A ``surprising'' agreement}

The photodetachment process from a single open-shell anion is written as
\begin{equation}
\label{eq:pd_process}
\fl
\mbox{X}^- [n_il_i^{N}\;  (S_i L_i) J_i] \; \; + \; \; (\hbar \omega)  
\rightarrow \mbox{X} [n_il_i^{N-1} \; (S_a L_a ) J_a] \; \; + \; \;  e^- [(sl_c)j_c]
\end{equation}
where the $i,a$ and $c$ indices refer to the negative {\it i}on, the neutral {\it a}tom and the {\it c}ontinuum electron, respectively.

\subsection{The ``standard'' approach}
Assuming pure LS coupling and using the Wigner law~\cite{Wig:48a} in the vicinity of the photodetachment threshold for setting the quantum numbers of the ejected electron $(l_c=0, j_c=s=1/2)$, Blondel {\it et al} \cite{Bloetal:06a} derived the relative intensities of the fine structure components for the detachment of a $p$-electron from the ``standard'' Wigner-Racah algebra \cite{Rac:42b,Jud:98a} in the electric dipole approximation :
\begin{equation}
\label{eq:stand_approach_1}
\fl
\mathcal{I}(J_a,J_i)=
\sum_J
[\,J_a,J_i,J\,]
\sixj{L_a}{J}{S_i}{s}{S_a}{J_a}^2
\sixj{J}{1}{J_i}{L_i}{S_i}{L_a}^2 
\end{equation}
with the abbreviated notation
\[
[ j_1,j_2,\ldots ] \equiv (2j_1 + 1)(2j_2 + 1) \ldots
\]

\subsection{The fractional parentage formula}
Describing the photodetachment as a direct one-electron process in which the laser photon ``annihilates'' an electron of angular momentum  $l_i$  to promote it into the continuum, the relative intensities can be calculated from the formula of Engelking and Lineberger~\cite{EngLin:79a}
\begin{equation}
\label{eq:fract_parentage_1}
\fl
\mathcal{I}(J_a,J_i)=
\sum_{j_i}
[\,J_a,J_i,j_i\,]
\ninej{S_a}{L_a}{J_a}{s}{l_i}{j_i}{S_i}{L_i}{J_i}^2 ,
\end{equation}
using the fractional parentage approach of Cox~\cite{Coxetal:72a,Cox:75a}. Expression \eref{eq:fract_parentage_1} is hereafter referred to as the ``Cox-Engelking-Lineberger'' fractional parentage formula.

\subsection{The S$^-$/S relative branching ratios of the fine-structure thresholds}
The relative branching ratios of the fine-structure thresholds for the $s$-wave photodetachment of $^{32}$S$^-$
\begin{equation}
\label{eq:Sulfur_pd_process}
\fl
\mbox{S}^- \; [3p^5\;  ^2P^o_{J_i} ] \; \; + \; \; (\hbar \omega)  
\rightarrow \mbox{S} \; [3p^4 \; ^3P_{J_a}] \; \; + \; \;  e^- [(l_c=0 ; j_c = 1/2)] .
\end{equation}
are reported in Table~\ref{tab:table1}, according to Blondel {\it et al}~\cite{Bloetal:06a}.
As observed by these authors, the two formalisms based on 
equations \eref{eq:stand_approach_1} and \eref{eq:fract_parentage_1} 
yield identical results. A similar agreement between the two approaches, 
that was presented as a ``surprise'' at first sight of the rather different 
expressions, was found in the study of the hyperfine structure relative intensities 
of photodetachment of $^{17}$O$^-$~\cite{Bloetal:01a}.


\begin{table}[bthp]
\caption{Relative branching ratios of the fine-structure thresholds.%
\label{tab:table1}}
\begin{center}
\begin{tabular}{@{}cccrcc}
\br
&& \\
$J_i$(S$^{-}$)   &  $J_a$(S)   &  $\mathcal{I}(J_a,J_i)$ \\
&& \\
\mr
&&& \\
1/2   &  0      &  4/54  \\
      &  1      &  9/54  \\
      &  2      &  5/54  \\
&&& \\
3/2   &  0      &  2/54  \\
      &  1      &  9/54  \\
      &  2      &  25/54 \\
&&& \\
\br
\end{tabular}
\end{center}
\end{table}


\subsection{An interesting angular momentum algebra relation}
The agreement between the numerical results obtained from equations \eref{eq:stand_approach_1} 
and \eref{eq:fract_parentage_1} is not limited to the above quantum number values and 
can not be accidental. We found, using a graphical approach~\cite{Bri:71a,LinMor:82a,Varetal:88a} 
presented in Appendix~A, an interesting {\it general} angular momentum relation
\begin{equation}
\label{eq:from9jto6j_gen}
\fl
\sum_j [j]
\ninej{j_1}{j_2}{j_3}{j_4}{j_5}{j}{j_6}{j_7}{j_8}^2
=
\sum_{j'} [j']
\sixj{j_2}{j_6}{j'}{j_4}{j_3}{j_1}^2
\sixj{j_2}{j_6}{j'}{j_8}{j_5}{j_7}^2 ,
\end{equation}
that, as shown in Appendix, is a special case of equation~(33)/sect.12.2 of Varshalovich {\it al.}~\cite{Varetal:88a}.

Applied in our context, relation \eref{eq:from9jto6j_gen} gives
\begin{equation}
\label{eq:from9jto6j_spec}
\fl
\sum_\alpha [\alpha]
\ninej{S_a}{L_a}{J_a}{s}{l_i}{\alpha}{S_i}{L_i}{J_i}^2
=
\sum_\beta [\beta]
\sixj{L_a}{\beta}{S_i}{s}{S_a}{J_a}^2
\sixj{\beta}{l_i}{J_i}{L_i}{S_i}{L_a}^2 .
\end{equation}
To the knowledge of the authors, relation \eref{eq:from9jto6j_gen}  cannot be found 
as such  in the current literature.

The link between \eref{eq:from9jto6j_spec} with the ``standard'' and 
fractional parentage formulae is established as follows:
\begin{enumerate}
\item
In the l.h.s of \eref{eq:from9jto6j_spec}, $\alpha$ plays the role 
in the fractional parentage formalism (equation~\eref{eq:fract_parentage_1}) 
of the possible $j$-values of the extracted electron in the negative ion, 
ie.  $\alpha = j_i = l_i \pm 1/2$.
\item
In the r.h.s of \eref{eq:from9jto6j_spec}, $\beta$ plays the role in 
the standard approach (equation~\eref{eq:stand_approach_1}) of the total angular 
momentum $J$ of the composite system (neutral atom + electron), ie. $ \beta = (J_a s)J = (J_aj_c)J$.
\end{enumerate}
Note that the entry ``1'' in the middle of the upper line of the second $6j$-symbol 
of the standard formula \eref{eq:stand_approach_1} corresponds to the rank one 
of the electric dipole (E1) transition operator. It appears in our 
relation \eref{eq:from9jto6j_spec} as the angular momentum value of the shell 
loosing one electron in the photodetachment process, restricting the above analysis 
to the photodetachment from a $p$-shell. However, this restriction is not too 
serious since this is precisely what Blondel {\it et al}~\cite{Bloetal:06a} needed 
in their ``standard'' approach for generating the $s$-outgoing electron wave as 
the dominant channel from Wigner's threshold law.

\section{Pan and Starace's analysis}

Pan and Starace \cite{PanSta:93a} parametrized the relative photoionization 
and photodetachment cross sections for fine structure transitions, starting from
\begin{equation}
\label{eq:PS_cross_section}
\fl
\sigma(J_a,J_i) = \frac{4\pi^2\omega}{c[J_i]} \sum_{M_i M l_c j_c J } 
\big\vert\, 
\mbox{\boldmath $ \hat{ \epsilon } $} \cdot
\bra< 
(S_a L_a ) J_a, (sl_c)j_c, J M - |
{\bf D}
\ket| 
(S_iL_i)J_iM_i > 
\,\big\vert^2 ,
\end{equation}
where $ {\bf D} \equiv \sum_{k=1}^N {\bf r} _k $ is the electric dipole operator 
and \mbox{\boldmath $ \hat{ \epsilon } $} is the polarization vector of the incident 
light of frequency $\omega$. The minus sign appearing in the bra indicates that 
the final state wave functions satisfy incoming-wave boundary conditions \cite{Sta:82a}. 
The final state of the composite system (neutral atom + electron) is characterized 
by the total angular momentum $J$ using the $(J_a j_c)J$ coupling, 
where $(s l_c) j_c$ results from the coupling of the spin ($s=1/2$) and the 
outgoing partial wave associated to the continuum photoelectron.
For the photodetachment process \eref{eq:pd_process}, they got the following 
{\it general} result\footnote{We observed that the square of the $3j$-symbol is missing in~(7) 
of Pan and Starace~\cite{PanSta:93a}. This has been confirmed by Starace~\cite{Sta:07a}.}
\begin{eqnarray}
\label{eq:PS_eq7}
\fl
\lefteqn{
\sigma(J_a,J_i)  = \frac{4\pi^2\omega}{3c} \; [J_a, S_i, L_i,l_i] \; N \; 
\big(\,S_aL_a,l_i \,\big\vert\big\}\,S_iL_i\,\big) ^2 }
\\
&  \times &
\sum_{l_c} [l_c]
\sum_{L}
\sum_{L'}
[L,L']
\threej{l_c}{1}{l_i}{0}{0}{0}^2
\big(\, \epsilon l_c 
\,\vert\, r \,\vert\,
n_i l_i \,\big)_{L}
\big(\, \epsilon l_c 
\,\vert\, r \,\vert\,
n_i l_i \,\big)_{L'}
 \exp{i(\phi_{\epsilon l_c}^L 
        - \phi_{\epsilon l_c}^{L'} )}
\nonumber
\\
& \times &
\sixj{l_c}{l_i}{1}{L_i}{L}{L_a}
\sixj{l_c}{l_i}{1}{L_i}{L'}{L_a}
\left[
\begin{array}{cccccccccc}
L_a & ~ & S_a & ~ & S_i & ~ & L_i & ~ & L & ~ \\
 ~ & J_a & ~ & 1/2 & ~ & J_i & ~ & 1 & ~ & l_c \\
L_a & ~ & S_a & ~ & S_i & ~ & L_i & ~ & L' & ~ \\
\end{array} \right]    ,  \nonumber
\end{eqnarray}
where  $\big(\, \epsilon l_c \,\vert\, r \,\vert\,n_i l_i \,\big)_{L}$ 
is the one-electron radial E1 matrix element depending on the $LS$ 
quantum numbers of the transition, $\phi_{\epsilon l_c}^L $ is the 
phase shift of the photoelectron with respect to a plane wave \cite{Sta:82a} 
and  $L$ ($L'$) appears as the angular momentum of the composite 
system [neutral atom ($L_a$) + electron ($l_c$)]. The last symbol with 
the 15 entries is a $15j$-symbol of the  second kind~\cite{MatBra:89a,Varetal:88a}. 

\subsection{The ``standard'' formula: a special case of \eref{eq:PS_eq7}}
Considering the case of the photodetachment of an open $p$-subshell electron 
and setting $l_c=0$ according to Wigner's threshold law, allows for a complete 
separation of dynamical and geometric factors and reduces \eref{eq:PS_eq7} to:
\newpage
\begin{eqnarray}
\label{eq:PS_eq12}
\fl
\lefteqn{
\sigma_{l=0}(J_a,J_i)
=
\frac{4\pi^2\omega}{3c} [J_a, S_i, L_i] \; N \; 
\big(p^{N-1} \,S_aL_a,p \,\big\vert\big\}\,p^N S_iL_i\,\big) ^2 }
\nonumber
\\
&  \times & 
\big(\, \epsilon s 
\,\vert\, r \,\vert\,
n_i l_i \,\big)_{L=L_a}^2 \delta_{l_i,1}
\sum_J
[\,J\,]
\sixj{L_a}{S_i}{J}{s}{J_a}{S_a}^2
\sixj{L_a}{S_i}{J}{J_i}{1}{L_i}^2 .
\end{eqnarray}
In this expression, one recognizes the summation over the product of the 
two squared $6j$-symbols appearing in the standard approach formula~\eref{eq:stand_approach_1}. 
As pointed out in the introduction, the fact that Pan and Starace's analysis \cite{PanSta:93a}  
integrated this result as a special case of \eref{eq:PS_eq7}, escaped to the attention 
of the authors of publications~\cite{Bloetal:06a} and \cite{Bloetal:01a}. 

\subsection{The term-independent approximation}
Pan and Starace~\cite{PanSta:93a} have also shown that, if the radial matrix elements  
are assumed independent of the angular momenta $L$ and $L'$ (the so-called 
{\it term-independent}~(TI) approximation), the partial photoionization cross 
section \eref{eq:PS_eq7} reduces to
\begin{eqnarray}
\label{eq:PS_eq13}
\fl
\lefteqn{
\sigma^{TI}(J_a,J_i)
=
\frac{4\pi^2\omega}{3c}
 \sum_{l_c} [l_c] \threej{l_c}{1}{l_i}{0}{0}{0}^2
\big(\, \epsilon l_c 
\,\vert\, r \,\vert\,
n_i l_i \,\big)^2 \;
[J_a, S_i, L_i] \; }
\\ & \times &
N \;
\big(\,S_aL_a,l_i \,\big\vert\big\}\,S_iL_i\,\big) ^2 \; 
\sum_J
[\,J\,]
\sixj{L_a}{S_i}{J}{s}{J_a}{S_a}^2
\sixj{L_a}{S_i}{J}{J_i}{1}{L_i}^2 . \nonumber
\end{eqnarray}

\subsection{Linking Pan and Starace TI cross section with previous works}
Pan and Starace \cite{PanSta:93a} were linking the partial photodetachment 
cross section derived in the term-independent approximation with all previous 
results \cite{Cox:75a,EngLin:79a,Schetal:80a,BerGoo:79a,GooBer:91a} through 
the following short statement: \\

``{\it Equation \eref{eq:PS_eq13}\footnote{numbered as (13) in the original reference~\cite{PanSta:93a}.} 
is equivalent to the single-configuration, 
$LS$-coupling, term-independent results of others}'', \\

\noindent referring to a brief footnote commenting the existence of some 
relations between ``{\it the sum over a squared $9j$-coefficient and an 
alternative way of representing the same $12j$-coefficient that we represent 
as a sum over a product of squared $6j$-coefficients}''. \\

\noindent The relation behind this footnote is nothing else than the angular 
momentum algebra relation  \eref{eq:from9jto6j_gen} demonstrated in Appendix~A.

\section{Using the irreducible tensorial expression of second quantization operators}

Using the spherical components of the photon polarization vector and of the 
electric dipole moment, the scalar product appearing in \eref{eq:PS_cross_section} is written as
\cite{Cow:81a}
\begin{equation}
\label{eq:dipole_mom}
\fl
 \mbox{\boldmath $ \hat{ \epsilon } $} \cdot {\bf D} = \sum_{q=-1}^{+1} \epsilon^{(1)}_{-q}  D^{(1)}_{q} 
  = \sum_{q=-1}^{+1} \epsilon^{(1)\ast}_q  D^{(1)}_{q} \,.
\end{equation}
Applying the Wigner-Eckart theorem and using the $3j$-symbol orthogonality, one easily finds~\cite{Sob:72a}
\begin{equation}
\label{eq:partial_cross_sect}
\fl
\sum_{M_i M }  
\big\vert\, 
\mbox{\boldmath $ \hat{ \epsilon } $} \cdot
\bra< \gamma J M -   |
{\bf D}
\ket|  \gamma_i J_iM_i > 
\,\big\vert^2
= \frac{1}{3} \big\vert\, 
\bra< \gamma J - | 
\red| D^{(1)} |
\ket| \gamma_i J_i  >
\,\big\vert^2
\, ,
\end{equation}
where the minus sign in the bra indicates that we refer to the wave function satisfying the incoming-wave boundary conditions~\cite{Sta:82a}.
The partial cross section \eref{eq:PS_cross_section} then reduces to
\begin{equation}
\label{eq:partial_cross_sect_2}
\fl
\sigma(J_a,J_i)  =
\frac{4\pi^2\omega}{3c[J_i]} \sum_{l_c }
\mathcal{D}^{l_c}(J_a,J_i)
\,,
\end{equation}
with
\begin{equation}
\label{eq:D_ronde_lc}
\fl
\mathcal{D}^{l_c}(J_a,J_i)
\equiv
\sum_{j_c J}
\big\vert\,
\bra< (S_a L_a ) J_a, (s,l_c)j_c, J - |
\red| 
D^{(1)} |
\ket| (S_iL_i)J_i >
\,\big\vert^2
\,.
\end{equation}

\subsection{Second quantized form of transition operators}

In the second quantization formalism~\cite{Jud:67a,Jud:96a}, any one-body operator $F = \sum_i f_i$ takes the form
\begin{equation}
\label{eq:one-body_sq}
\fl
F = \sum_{\xi,\eta}  a^{\dag}_\xi \; \langle \xi \vert f \vert \eta \rangle \; a_\eta \; .
\end{equation}
The creation $a^{\dag}_{\sigma}$  operators, where $\sigma$ stands for
$(n l m_s m_l)$, form the components of a double tensor $\acreate{(s l)}$ of 
rank $s$ with respect to spin and rank $l$ with respect to orbit~\cite{Jud:67a}. 
Similarly, a double tensor can be created from the collection of annihilation operators but 
a phase factor must be introduced~\cite{Jud:67a} for defining the components $\tilde{a}_\sigma$ 
\[ \tilde{a}_{n l m_s m_l} = (-1)^{s+l-m_s-m_l} a_{nl-m_s -m_l} \]
that form the double tensor ${\ensuremath{\t{a}{(s l)}}}$.
It becomes then possible to build the coupled tensors~\cite{Cow:81a}
\begin{equation}
\label{eq:coupled_tensors_sq}
\fl
\Big[
\acreate{(s l)}
\times
\aanni{(s l')}
\Big]^{(\kappa k)}_{\pi q} = 
\sum_{\xi,\eta} \;
(s m_{s_\xi} s m_{s_\eta} \vert s s \kappa \pi )
(l m_{l_\xi} l' m_{l_\eta} \vert l l' k q ) \; 
a_\xi^\dag \tilde{a}_\eta \, .
\end{equation}
Using the atomic shell theory~\cite{Rud:97a}, the second quantized form of 
the one-electron operator \eref{eq:one-body_sq} is written in the 
following $(SL)J$-coupling tensorial form~\cite{GaiRud:96a}:
\begin{eqnarray}
\label{eq:equiv_GaiRud_eq3}
\fl
F = \sum_{n_il_i, n_jl_j} 
[K_S,K_L]^\mhalf
(n_i s l_i \Vert f^{K_S K_L} \Vert n_j s l_j )
\Big[
\acreate{(s l_i)}
\times
\aanni{(s l_j)}
\Big]^{(K_S K_L)K_J}_{M_J}
\, ,
\end{eqnarray}
where $K_S, K_L$ and $K_J$ specify the rank with respect to spin, orbit and total angular momentum, respectively, and 
where $(n_i s l_i \Vert f^{K_S K_L} \Vert n_j s l_j )$ is the appropriate one-electron reduced matrix element.
In the single-configuration picture, one can pick up from this double sum over the active shells the specific term 
inducing the desired one-electron jump $i \leftarrow j$, ie.
\begin{eqnarray}
\label{eq:op_LSJ_ij_gen}
\fl
\t{T}{(K_SK_L)K_J}_{ M_J} (i \leftarrow j)
&
=
&
[K_S,K_L]^\mhalf
(n_i s l_i \Vert f^{K_S K_L} \Vert n_j s l_j )
\Big[
\acreate{(s l_i)}
\times
\aanni{(s l_j)}
\Big]^{(K_S K_L)K_J}_{M_J}
\, .
\end{eqnarray}
For the photodetachment process~\eref{eq:pd_process} described in the electric dipole approximation, 
the transition operator appearing in \eref{eq:D_ronde_lc} has the tensorial structure $(K_SK_L)K_J=(01)1$ and  
is written as
\begin{eqnarray}
\label{eq:op_LSJ_ij_E1}
\fl
\t{T}{(01)1}_Q (\epsilon l_c \leftarrow n_i l_i)
&
=
&
[1]^\mhalf \; 
t(\epsilon s l_c, n_i s l_i) \; 
\Big[
\acreate{(s l_c)}
\times
\aanni{(s l_i)}
\Big]^{(01)1}_Q \hspace*{1cm} ; \hspace*{1cm} Q=0, \pm 1
\,, 
\end{eqnarray}
where $t(\epsilon s l_c, n_i s l_i)$ stands for the one-electron E1 reduced matrix element 
\begin{eqnarray}
t(\epsilon s l_c, n_i s l_i)
&
=
&
\bra< \epsilon s l_c |
\red| \t{t}{(01)} |
\ket| n_i s l_i >
\equiv
\bra< \epsilon s l_c |
\red| \t{S}{(0)} \t{C}{(1)} r |
\ket| n_i s l_i >
\nonumber
\,. 
\end{eqnarray}
As suggested by Pan and Starace~\cite{PanSta:93a} and by equation~\eref{eq:PS_eq7}, one needs to integrate in its expression a phase 
factor for the incoming-wave boundary conditions~\cite{Sta:82a}, together with an explicit\footnote{also implicitly containing
the quantum numbers $S, L_a$ and $ S_a$.} subscript $L$   for discussing a possible 
term-dependency of the radial matrix element~\cite{Diletal:75a}:
\begin{eqnarray}
\label{eq:red_sing_matrix}
t_L(\epsilon s l_c, n_i s l_i)
&
=
&
\bra< \epsilon s l_c |
\red| \t{t}{(01)} |
\ket| n_i s l_i >_L
=
\bra< \epsilon s l_c |
\red| \t{S}{(0)} \t{C}{(1)} r |
\ket| n_i s l_i >_L
\nonumber
\\
&
=
&
(-1)^{l_c}
[s,l_c,l_i]^\half
\threej{l_c}{1}{l_i}{0}{0}{0}
\big(\, \epsilon l_c 
\,\vert\, r \,\vert\,
n_i l_i \,\big)_L \exp{i(\phi_{\epsilon l_c}^L)}
\,. 
\end{eqnarray}

\subsection{In $(SL)J$-coupling}
The $\mathcal{D}^{l_c}(J_a,J_i)$  contribution \eref{eq:D_ronde_lc} to the partial cross section \eref{eq:partial_cross_sect_2} 
is written as
\begin{eqnarray}
\label{eq:crossection_LS_start_1}
\fl
\mathcal{D}^{l_c}(J_a,J_i)
=
&
\sum_{J j_c}
\Big\vert\,
\bra< (S_aL_a)J_a, (s l_c)j_c,J|
\red| \t{T}{(01)1}_{\epsilon l_c \leftarrow n_i l_i} |
\ket| (S_iL_i)J_i >
\,\Big\vert^2 \, .
\end{eqnarray}
To evaluate the matrix element, one first recouples the final combined system, transforming the bra 
from $jj$- to $SL$-coupling \cite{Cow:81a,Kar:96a}
\begin{eqnarray}
\label{eq:LS_final}
\fl
\bra< (S_aL_a)J_a,(sl_c)j_c,J| 
&
=
&
\sum_{SL}
[S,L,J_a,j_c]^\half
\ninej{S_a}{s}{S}{L_a}{l_c}{L}{J_a}{j_c}{J}
\bra<(S_as)S,(L_al_c)L,J|
\,.
\end{eqnarray}
Using \eref{eq:op_LSJ_ij_E1} and \eref{eq:LS_final}, \eref{eq:crossection_LS_start_1} becomes
\begin{eqnarray}
\label{eq:crossection_LS_start_2}
\fl
\mathcal{D}^{l_c}(J_a,J_i)
=
&
\sum_{J j_c}
\Big\vert\,
\sum_{SL}
t_L(\epsilon s l_c, n_i s l_i)
[S,L,J_a,j_c]^\half
[1]^\mhalf
\ninej{S_a}{s}{S}{L_a}{l_c}{L}{J_a}{j_c}{J}
\nonumber
\\
& \times
\bra<(S_as)S,(L_al_c)L,J|
\red|
\Big[
\acreate{(s l_c)}
\times
\aanni{(s l_i)}
\Big]^{(01)1}
|
\ket| (S_iL_i)J_i>
\,\Big\vert^2
\,.
\end{eqnarray}
The $J$-dependency within the reduced matrix element of the $(SL)J$ coupled creation and annihilation tensor product is 
extracted using
\newpage
\begin{eqnarray}
\label{eq:rme_cre_ann}
\fl
\lefteqn{
\bra<(S_as)S,(L_al_c)L,J|
\red|
\Big[
\acreate{(s l_c)}
\times
\aanni{(s l_i)}
\Big]^{(01)1}
|
\ket| (S_iL_i)J_i>
\, 
= } \\ 
& &
[J,1,J_i]^\half \; 
\ninej{S}{L}{J}{S_i}{L_i}{J_i}{0}{1}{1}
\bra<(S_as)S,(L_al_c)L|
\red|
\Big[
\acreate{(s l_c)}
\times
\aanni{(s l_i)}
\Big]^{(01)}
|
\ket| S_iL_i>
\, . \nonumber
\end{eqnarray}
Thanks to the zero entry,  the $9j$-symbol simplifies to
\begin{eqnarray}
\ninej{S}{L}{J}{S_i}{L_i}{J_i}{0}{1}{1}
& = &
(-1)^{L_i+J+S_i+1}
[1,S_i]^\mhalf
\sixj{J_i}{L_i}{S_i}{L}{J}{1}
\,.
\end{eqnarray}
After inserting explicitly the empty continuum space into the bra describing the ion state,
\eref{eq:crossection_LS_start_2} becomes
\begin{eqnarray}
\label{eq:crossection_LS_1}
\fl
\mathcal{D}^{l_c}(J_a,J_i)
=
&
\sum_{J j_c}
\Big\vert\,
\sum_{SL}
t_L(\epsilon s l_c, n_i s l_i)
[S,L,J_a,j_c,J,J_i]^\half
[1,S_i]^\mhalf
\nonumber
\\
& \times
(-1)^{L_i+J+S_i+1}
\ninej{S_a}{s}{S}{L_a}{l_c}{L}{J_a}{j_c}{J}
\sixj{J_i}{L_i}{S_i}{L}{J}{1}
\nonumber
\\
& \times
\bra<(S_as)S,(L_al_c)L|
\red|
\Big[
\acreate{(s l_c)}
\times
\aanni{(s l_i)}
\Big]^{(01)}
|
\ket| (S_i0)S_i,(L_i0)L_i>
\,\Big\vert^2
\,.
\end{eqnarray}
The matrix element of the tensor product of creation and annihilation operators is expressed in terms of submatrix elements involving the individual operators by introducing a summation over a complete set of intermediate states $S'L'$ \cite{Cow:81a,Kar:96a},
\begin{eqnarray}
\label{eq:crossection_LS_3}
\fl
\mathcal{D}^{l_c}(J_a,J_i)
=
&
\sum_{J j_c}
\Big\vert\,
\sum_{SL}
t_L(\epsilon s l_c, n_i s l_i)
[S,L,J_a,j_c,J,J_i]^\half
[1,S_i]^\mhalf
\nonumber
\\
& \times
(-1)^{L_i+J+S_i+1}
\ninej{S_a}{s}{S}{L_a}{l_c}{L}{J_a}{j_c}{J}
\sixj{J_i}{L_i}{S_i}{L}{J}{1}
\nonumber
\\
& \times
(-1)^{S+S_i+L+1}
[1]^\half
\sum_{S'L'}
\sixj{s}{s}{0}{S_i}{S}{S'}
\sixj{l_c}{l_i}{1}{L_i}{L}{L'}
\nonumber
\\
& \times
\bra<(S_as)S,(L_al_c)L|
\red|
\acreate{(s l_c)}
|
\ket|(S'0)S',(L'0)L'>
\nonumber
\\
& \times
\bra<(S'0)S',(L'0)L'|
\red|
\aanni{(s l_i)}
|
\ket| (S_i0)S_i,(L_i0)L_i>
\,\Big\vert^2
\,.
\end{eqnarray}
The reduction of the $6j$-symbol 
\begin{eqnarray}
\sixj{s}{s}{0}{S_i}{S}{S'}
&=&
\delta(s S S') (-1)^{s+S+S'}
[s,S]^\mhalf
\del(S_i,S)
\end{eqnarray}
simplifies the summation over $S$, thanks to the Kronecker delta\footnote{The $\delta(ijk)$ notation represents $+1$ if the triangle relations are satisfied and $0$ otherwise.}. 
After realizing that the creation operator acts only on the continuum space,
the first reduced matrix element appearing in \eref{eq:crossection_LS_3} is evaluated 
by using the uncoupling formula for reduced matrix elements \cite{Cow:81a,Kar:96a}
\begin{eqnarray}
\label{eq:uncplcreation}
\fl
\lefteqn{
\bra<(S_as)S,(L_al_c)L|
\red|
\acreate{(s l_c)}
|
\ket|(S'0)S',(L'0)L'>
\, 
 } \nonumber \\ 
& &
=  \delta(S_a,S')  \delta(L_a,L')  (-1)^{S_a + S + s + L_a + L + l_c} [S,S',L,L']^\half 
\nonumber \\
& &
\times 
\sixj{S_a}{s}{S}{s}{S'}{0} \sixj{L_a}{l_c}{L}{l_c}{L'}{0}
\bra<sl_c|
\red|
\acreate{(s l_c)}
|
\ket|00>
\,.
\end{eqnarray}
Using the reduced matrix element of the creation operator
\begin{eqnarray}
\label{eq:creation_rme}
\bra<sl_c|
\red|
\acreate{(s l_c)}
|
\ket|00>
&=& 
-[s,l_c]^\half \, ,
\end{eqnarray}
in agreement with the $N=1$ limit case of Judd's expression~\cite{Jud:67a})
\begin{eqnarray}
\label{eq:Judd_creation_rme}
\bra<\psi|
\red|
\acreate{\ }
|
\ket|\overline{\psi}>
&=& (-1)^N \{ N [S,L] \}^\half ( \psi \{ | \overline{\psi})  \, ,
\end{eqnarray}
\eref{eq:uncplcreation}  becomes
\begin{eqnarray}
\label{eq:uncplcreation_2}
\fl
\lefteqn{
\bra<(S_as)S,(L_al_c)L|
\red|
\acreate{(s l_c)}
|
\ket|(S'0)S',(L'0)L'>
\, 
 } \nonumber \\ 
& &
= - [S,L]^\half \; \delta(S_a s S) \delta(L_a L_c L) \; \delta(S_a,S')  \delta(L_a,L') 
\,.
\end{eqnarray}

The second reduced matrix element appearing in \eref{eq:crossection_LS_3} is worked out similarly 
for the annihilation operator acting in the $n_i l_i$ shell space 
\begin{eqnarray}
\label{eq:uncplannihilation}
\fl
\lefteqn{
\bra<(S'0)S',(L'0)L'|
\red|
\aanni{(s l_i)}
|
\ket| (S_i0)S_i,(L_i0)L_i>
\, 
 } \nonumber \\ 
& &
=   (-1)^{S' + S_i + s + L' + L_i + l_i} [S',S_i,L',L_i]^\half 
\nonumber \\
& &
\times 
\sixj{S'}{S_i}{s}{S_i}{S'}{0} \sixj{L'}{L_i}{l_i}{L_i}{L'}{0} 
\bra<S'L'|
\red|
\aanni{(s l_i)}
|
\ket| S_iL_i>
\, .
\end{eqnarray}
Using the annihilation operator reduced matrix element (see equation~(32) of~\cite{Jud:67a})
\begin{eqnarray}
\label{eq:annihilation_rme}
\bra<S'L'|
\red|
\aanni{(s l_i)}
|
\ket| S_iL_i>
&
=
&
\sqrt{N_i} 
(-1)^{N_i+S'-s-S_i+L'-l_i-L_i}
\nonumber \\
&&\times
\; [S_i,L_i]^\half \; 
\big(\,S'L',l_i \,\big\vert\big\}\,S_iL_i\,\big) \, ,
\end{eqnarray}
\eref{eq:uncplannihilation}  becomes
\begin{eqnarray}
\label{eq:uncplannihilation_2}
\fl
\lefteqn{
\bra<(S'0)S',(L'0)L'|
\red|
\aanni{(s l_i)}
|
\ket| (S_i0)S_i,(L_i0)L_i>
\, 
 }  \\ 
& &
= \delta(S' S_i s) \delta(L' L_i l_i) \; \sqrt{N_i} 
(-1)^{N_i+S'-s-S_i+L'-l_i-L_i} 
\; [S_i,L_i]^\half \; 
\big(\,S'L',l_i \,\big\vert\big\}\,S_iL_i\,\big)
\,. \nonumber
\end{eqnarray}
Combining equations \eref{eq:crossection_LS_3}, \eref{eq:uncplcreation_2} and \eref{eq:uncplannihilation_2}, and taking $N_i = N$ according to \eref{eq:pd_process}, the summations over $S,S'$ and $L'$ are reduced to give
\begin{eqnarray}
\fl
\mathcal{D}^{l_c}(J_a,J_i)
=
&
\sum_{J j_c}
\Big\vert\,
(-1)^{J + N + S_i + L_a -l_i} \; 
\sqrt{N/2} \;
\big(\,S_aL_a,l_i \,\big\vert\big\}\,S_iL_i\,\big) \;
[J_a,j_c,J,J_i, S_i, L_i]^\half
\nonumber
\\
& \times
\sum_{L} t_L(\epsilon s l_c, n_i s l_i)
(-1)^L [L] 
 \nonumber
\\
& \times
\ninej{S_a}{s}{S_i}{L_a}{l_c}{L}{J_a}{j_c}{J}
\sixj{J_i}{L_i}{S_i}{L}{J}{1}
\sixj{l_c}{l_i}{1}{L_i}{L}{L_a}
\,\Big\vert^2
\nonumber
\, ,
\end{eqnarray}
\newpage
\begin{eqnarray}
\label{eq:crossection_LS_4}
\fl
\mathcal{D}^{l_c}(J_a,J_i)
= &
\frac{N}{2} \; [J_a,J_i, S_i, L_i] \;
\vert
\big(\,S_aL_a,l_i \,\big\vert\big\}\,S_iL_i\,\big)
\vert ^2
\nonumber
\\
& \times
\sum_{J j_c} \; [j_c,J] \;
\sum_{L}
\sum_{L'}
t_L(\epsilon s l_c, n_i s l_i)
t_{L'}^*(\epsilon s l_c, n_i s l_i)
(-1)^{L+L'} [L] [L']
\nonumber
\\
& \times
\ninej{S_a}{s}{S_i}{L_a}{l_c}{L}{J_a}{j_c}{J}
\ninej{S_a}{s}{S_i}{L_a}{l_c}{L'}{J_a}{j_c}{J}
\sixj{J_i}{L_i}{S_i}{L}{J}{1}
\sixj{J_i}{L_i}{S_i}{L'}{J}{1}
\nonumber
\\
& \times
\sixj{l_c}{l_i}{1}{L_i}{L}{L_a}
\sixj{l_c}{l_i}{1}{L_i}{L'}{L_a}
\, .
\end{eqnarray}
\noindent
Using graphical techniques~\cite{Bri:71a,LinMor:82a,Varetal:88a,MatBra:89a}, \eref{eq:crossection_LS_4} is finally rewritten as
\begin{eqnarray}
\label{eq:crossection_LS_5}
\fl
\mathcal{D}^{l_c}(J_a,J_i)
 & = 
\frac{N}{2} \; [J_a,J_i, S_i, L_i] \;
\vert
\big(\,S_aL_a,l_i \,\big\vert\big\}\,S_iL_i\,\big)
\vert ^2 \\
\fl
&
\times
\sum_{L}
\sum_{L'}
t_L(\epsilon s l_c, n_i s l_i)
t_{L'}^*(\epsilon s l_c, n_i s l_i)
[L,L'] \;
\sixj{l_c}{l_i}{1}{L_i}{L}{L_a}
\sixj{l_c}{l_i}{1}{L_i}{L'}{L_a}
\nonumber 
\\
\fl
&
\times
\left[
\begin{array}{cccccccccc}
L_a & ~ & S_a & ~ & S_i & ~ & L_i & ~ & L & ~ \\
 ~ & J_a & ~ & 1/2 & ~ & J_i & ~ & 1 & ~ & l_c \\
L_a & ~ & S_a & ~ & S_i & ~ & L_i & ~ & L' & ~ \\
\end{array} \nonumber
\right]
\,.
\end{eqnarray}
From the definition of the one-electron reduced matrix element~\eref{eq:red_sing_matrix}, 
the link with section~3 can be  done, in particular with Pan and Starace's general parametrization~\eref{eq:PS_eq7}, after using
\eref{eq:partial_cross_sect_2} for building the partial cross section. In other terms, one has reproduced  
Pan and Starace's cross section expression (see however the footnote on page~5), adopting the 
irreducible tensorial expression of the second quantized form of the electric dipole operator. 
The particular cases of the ``standard'' and the term-independent cross sections, discussed in section~3
(see equations~\eref{eq:PS_eq12} and \eref{eq:PS_eq13}, respectively), can obviously be derived from this common result.

\subsection{In $(jj)J$-coupling}
In the previous subsection, the calculation was performed in $(SL)J$-coupling.
The annihilation operator acted on the ion
and annihilated the electron $\ket|sl_i>$.  The creation operator acted on the
vacuum and created the photoelectron. This photoelectron was coupled to the
outgoing atom to intermediate $\ket|SL>$ states. These states were recoupled to
the final $\bra<(J_a,j_c)J|$ state. 
If one applies the term-independent
approximation, the summation over the intermediate states leads to the result
\eref{eq:PS_eq7} that is independent of the intermediate states.

In the present section, the term-independent approximation is used from the
very beginning.  The final state is obviously $(jj)J$-coupled. So is the
initial state if the continuum vacuum is added  
$\ket|(S_iL_i)J_i> = \ket|(S_iL_i)J_i,(00)0,J>$.  
The summation over intermediate states introduced in $(SJ)L$-coupling (see previous subsection)
is not needed if the spin-angular part of the operator,
that is the coupled tensorial product of the creation and annihilation
operator appearing in \eref{eq:op_LSJ_ij_gen}, is recoupled from $(K_SK_L)K_J$ to $(jj)J$ 
\begin{eqnarray}
\label{eq:LSJ_to_jjJ}
\t{T}{(K_SK_L)K_J}_{M_J} (i \leftarrow j)
=
[K_S,K_L]^\mhalf
(n_i s l_i \Vert f^{K_S K_L} \Vert n_j s l_j )
\Big[
\acreate{(s l_i)}
\times
\aanni{(s l_j)}
\Big]^{(K_S K_L)K_J}_{M_J}
\nonumber
\\
\fl
\quad
=
(n_i s l_i \Vert f^{K_S K_L} \Vert n s l_j )
\sum_{j_p j_q}
[j_p,j_q]^\half
\ninej{K_S}{K_L}{K_J}{s}{l_i}{j_p}{s}{l_j}{j_q}
\Big[
\acreate{(s l_i)}
\times
\aanni{(s l_j)}
\Big]^{(j_p j_q)K_J}_{M_J}
\,. 
\end{eqnarray}
The ranks $j_p$ and $j_q$ are used for the creation and annihilation operators, respectively.
This transformation is pure angular recoupling, without affecting the one-electron
matrix elements. In other words, the $(SL)J-(jj)J$ recoupling is performed
without invoking the full-relativistic approach\footnote{in which the
second-quantized creation operator to be used should be the operator
producing the 4-components Dirac spinor~\cite{Gra:07a}, i.e. $
a^{\dagger}_{n \kappa m} \vert 0 \rangle = \vert n \kappa m \rangle
$.}. The one-electron matrix elements are kept as the
non-relativistic, term-independent quantities used in the previous section.  
Setting the ranks to $K_S=0, K_L=1$ and $K_J=1$ for
the electric dipole photodetachment
process, with $Q=M_J=0,\pm1$, the operator \eref{eq:LSJ_to_jjJ} has the 
form
\begin{eqnarray}
\label{eq:op_LSJ_to_jjJ_E1}
\fl
\t{T}{(01)1}_Q (\epsilon l_c \leftarrow n_i l_i)
&
=
t(\epsilon s l_c, n_i s l_i)
\Big[
\acreate{(s l_c)}
\times
\aanni{(s l_i)}
\Big]^{(01)1}_Q
\nonumber
\\
\fl
&
=
t(\epsilon s l_c, n_i s l_i) \;
\sum_{j_p j_q}
[j_p,j_q]^\half
\ninej{0}{1}{1}{s}{l_c}{j_p}{s}{l_i}{j_q}
\Big[
\acreate{(s l_c)}
\times
\aanni{(s l_i)}
\Big]^{(j_p j_q)1}_Q
\,,
\end{eqnarray}
where the one-electron reduced matrix elements are the term-independent
form of 
\eref{eq:red_sing_matrix}:
\begin{eqnarray}
\label{eq:red_sing_matrix_TI}
\fl
t(\epsilon s l_c, n_i s l_i)
&
=
&
\bra< \epsilon s l_c |
\red| \t{t}{(01)} |
\ket| n_i s l_i >
= 
(-1)^{l_c}
[s,l_c,l_i]^\half
\threej{l_c}{1}{l_i}{0}{0}{0}
\big(\, \epsilon l_c 
\,\vert\, r \,\vert\,
n_i l_i \,\big)
\,.
\end{eqnarray}
Using  the following reduction 
\begin{eqnarray}
\label{eq:simplify_9j_jjJ}
\ninej{0}{1}{1}{s}{l_c}{j_p}{s}{l_i}{j_q}
&
=
&
(-1)^{j_p+l_i+1+s}
[1,s]^\mhalf
\sixj{j_q}{j_p}{1}{l_c}{l_i}{s}
\, ,
\end{eqnarray}
the transformed electric dipole operator simplifies to
\begin{eqnarray}
\label{eq:op_jjJ_dipole}
\fl
\t{T}{(01)1}_Q (\epsilon l_c \leftarrow n_i l_i)
=
&
t(\epsilon s l_c, n_i s l_i) \;
\sum_{j_p j_q}
[j_p,j_q]^\half
[1,s]^\mhalf
(-1)^{j_p+l_i+1+s}
\nonumber
\\
&\times
\sixj{j_q}{j_p}{1}{l_c}{l_i}{s}
\Big[
\acreate{(s l_c)}
\times
\aanni{(s l_i)}
\Big]^{(j_p j_q)1}_Q
\,,
\end{eqnarray}
that is used for expressing \eref{eq:crossection_LS_start_1} as
\begin{eqnarray}
\label{eq:Dcontrijj}
\fl
\mathcal{D}^{l_c}(J_a,J_i)
=
\sum_{J j_c}
\Big\vert\,
t(\epsilon s l_c, n_i s l_i) \; 
\sum_{j_p j_q} \; 
[j_p,j_q]^\half
[1,s]^\mhalf
(-1)^{j_p+l_i+1+s}
\sixj{j_q}{j_p}{1}{l_c}{l_i}{s}
\nonumber
\\
\times
\bra< (S_aL_a)J_a, (s l_c)j_c,J|
\red|
\Big[
\acreate{(s l_c)}
\times
\aanni{(s l_i)}
\Big]^{(j_p j_q)1}
|
\ket|(S_i L_i) J_i,(00)0, J_i >
\,\Big\vert^2
\,.
\end{eqnarray}
The annihilation operator in \eref{eq:Dcontrijj} acts between the ion and the
remaining atom while the creation operator acts on a different subset between the
continuum vacuum and the free electron.  
Using the decoupling formula~\cite{Cow:81a,Kar:96a},
the contributions of the different subspaces factorize:
\begin{eqnarray}
\label{eq:crossection_jj_1}
\fl
\bra< (S_aL_a)J_a, (s l_c)j_c,J|
\red|
\Big[
\acreate{(s l_c)}
\times
\aanni{(s l_i)}
\Big]^{(j_p j_q)1}
|
\ket|(S_i L_i) J_i,(00)0,J_i >
\nonumber
\\
\fl
\quad
= (-1)^{J_a + j_c -J} 
[\, J, 1, J_i \,]^\half
\ninej{j_c}{0}{j_p}{J_a}{J_i}{j_q}{J}{J_i}{1}
\bra< (sl_c)j_c |
\red|
\acreate{(s l_c)j_p}
|\ket| (00)0 >
\bra< (S_aL_a)J_a|
\red|
\aanni{(s l_i)j_q}
|\ket| (S_i L_i) J_i >
\nonumber
\\
\fl
\quad
=
(-1)^{J + J_i + 1}
[\,J,1\,]^\half
[\,j_c\,]^\mhalf
\delta(j_c,j_p)
\sixj{j_q}{J_a}{J_i}{J}{1}{j_c}
\nonumber
\\
\fl
\quad
\quad
\times
\bra< (sl_c)j_c |
\red|
\acreate{(s l_c)j_p}
|\ket| (00) 0 >
\;
\bra< (S_aL_a)J_a|
\red|
\aanni{(s l_i)j_q}
|\ket| (S_i L_i) J_i >
\,.
\end{eqnarray}
The reduced matrix elements of the annihilation and creation operator
are calculated by eliminating the J dependence as follows
\begin{eqnarray}
\label{eq:rme_annihil_1}
\fl
\bra< (S_aL_a)J_a|
\red|
\aanni{(s l_i)j_q}
|\ket| (S_i L_i) J_i >
=
[\,J_a,j_q,J_i\,]^\half
\ninej{S_a}{L_a}{J_a}{S_i}{L_i}{J_i}{s}{l_i}{j_q}
\bra< S_aL_a |
\red|
\aanni{(s l_i)}
|\ket| S_i L_i >
\end{eqnarray}
and
\begin{eqnarray}
\label{eq:rme_crea_1}
\fl
\bra< (sl_c)j_c |
\red|
\acreate{(s l_c)j_c}
|\ket| (00)0 >
&=&
[\,j_c\,]
\ninej{s}{l_c}{j_c}{0}{0}{0}{s}{l_c}{j_c}
\bra< sl_c |
\red|
\acreate{(s l_c)}
|\ket| 00 >
\nonumber
\\
&=&
[\,j_c\,]^\half
[\,l_c,s\,]^\mhalf
\bra< sl_c |
\red|
\acreate{(s l_c)}
|\ket| 00  > 
\delta(s l_c j_c)
\,,
\end{eqnarray}
Using equations \eref{eq:creation_rme} and \eref{eq:annihilation_rme}, 
the contribution \eref{eq:Dcontrijj} to the partial cross section becomes (setting $N_i=N$):
\begin{eqnarray}
\label{eq:crossection_jj_8}
\fl
\mathcal{D}^{l_c}(J_a,J_i)
=
\sum_{J j_c}
\Big\vert\,
(-1)^{N+ 1 + J + J_i + j_c + S_a - S_i + L_a - L_i} \; 
\sqrt{N/2} \; 
\big( 
S_aL_a,l_i
\big\vert 
\big\}
S_iL_i 
\big) \;
[j_c,J_a,S_i,L_i,J,J_i]^\half 
\nonumber
\\ 
\times \; 
\sum_{j_i} \;
t(\epsilon s l_c, n_i s l_i) \;
 [j_i]
\ninej{S_a}{L_a}{J_a}{S_i}{L_i}{J_i}{s}{l_i}{j_i}
\sixj{j_i}{j_c}{1}{l_c}{l_i}{s}
\sixj{j_i}{J_a}{J_i}{J}{1}{j_c}
\,\Big\vert^2
\,. 
\end{eqnarray}

Remembering that the one-electron matrix elements \eref{eq:red_sing_matrix} 
are $j_c$ and $j_i$ independent, one finally obtains a quadruple summation
\begin{eqnarray}
\label{eq:crossection_jj9}
\fl
\mathcal{D}^{l_c}(J_a,J_i)
&
=
&
\frac{N}{2} \;
\vert
\big( 
S_aL_a,l_i
\big\vert 
\big\}
S_iL_i 
\big) \;
\vert ^2 \; [J_a,S_i,L_i,J_i] 
\sum_{J} \;
[\,J\,] \; 
\vert
t(\epsilon s l_c, n_i s l_i) 
\vert ^2 \;
\nonumber
\\
&& \times
\sum_{j_c}
\sum_{j_i}
\sum_{j'_i}
[\,
j_c,
j_i,
j'_i\,]
\ninej{S_a}{L_a}{J_a}{S_i}{L_i}{J_i}{s}{l_i}{j_i}
\ninej{S_a}{L_a}{J_a}{S_i}{L_i}{J_i}{s}{l_i}{j'_i}
\nonumber
\\
&& \times
\sixj{j_i}{j_c}{1}{l_c}{l_i}{s}
\sixj{j'_i}{j_c}{1}{l_c}{l_i}{s}
\sixj{j_c}{J}{J_a}{J_i}{j_i}{1}
\sixj{j_c}{J}{J_a}{J_i}{j'_i}{1}
\,.
\end{eqnarray}
Moving the summation symbol over $j_c$ to the right and using 
equation~(33)/sect.12.2 of Varshalovich {\it al.}~\cite{Varetal:88a} (see also the Appendix)
\begin{eqnarray}
\label{eq:4_6j_to_2_9j_to_1_12j}
\fl
\sum_{j_c} \;
[j_c] \;
\sixj{j_i}{1}{j_c}{l_c}{s}{l_i}
\sixj{l_c}{s}{j_c}{j'_i}{1}{l_i}
\sixj{j'_i}{1}{j_c}{J}{J_a}{J_i}
\sixj{J}{J_a}{j_c}{j_i}{1}{J_i} 
&
& \nonumber \\
= \sum_{j_c} \; 
[j_c] \;
\ninej{j_i}{1}{j_c}{s}{l_i}{j'_i}{l_i}{l_c}{1}
\ninej{j_i}{1}{j_c}{J_a}{J_i}{j'_i}{J_i}{J}{1} 
= 
\left\{
\begin{array}{cccc}
-   &  j_i   &   s   &   l_i  \\
1   &  -     &  l_i  &   l_c  \\
J   &  J_i   &  -    &    1   \\
J_i &  J_a   &  j'_i &   -    
\end{array} \right\} \, ,
\end{eqnarray}
the summation over $j_c$ in \eref{eq:crossection_jj9} is incorporated into
the $12j$-symbol. 
A compact and elegant expression is obtained:
\begin{eqnarray}
\label{eq:crossection_jj_10}
\fl
\mathcal{D}^{l_c}(J_a,J_i)
=
\frac{N}{2} \;
\vert
\big( 
S_aL_a,l_i
\big\vert
\big\}
S_iL_i 
\big) \;
\vert ^2 \; [J_a,S_i,L_i,J_i] 
\sum_{J} \;
[J] \;
\vert
t(\epsilon s l_c, n_i s l_i) 
\vert ^2 \;
\nonumber
\\
\times
\sum_{j_i}
\sum_{j'_i}
[\,j_i,j'_i\,]
\ninej{S_a}{L_a}{J_a}{S_i}{L_i}{J_i}{s}{l_i}{j_i} \;
\ninej{S_a}{L_a}{J_a}{S_i}{L_i}{J_i}{s}{l_i}{j'_i} \;
\left\{
\begin{array}{cccc}
-   &  j_i   &   s   &   l_i  \\
1   &  -     &  l_i  &   l_c  \\
J   &  J_i   &  -    &    1   \\
J_i &  J_a   &  j'_i &   -    
\end{array} \right\} \, .
\end{eqnarray}
If one further assumes that the one-electron reduced matrix elements 
are $J$-independent, they can be moved to the left of the summation
symbol over $J$ to take advantage of the following identity
\begin{eqnarray}
\label{eq:summation_12j}
\fl
\sum_{J} \;
[J] \;
\left\{
\begin{array}{cccc}
-   &  j_i   &   s   &   l_i  \\
1   &  -     &  l_i  &   l_c  \\
J   &  J_i   &  -    &    1   \\
J_i &  J_a   &  j'_i &   -    
\end{array} \right\} 
= \del(j_i,j'_i) \; [l_i,j_i]^{-1} \;  \delta (l_i s j_i) \; \delta (j_i J_a J_i ) \;  \delta (l_i l_c 1)
\, ,
\end{eqnarray}
that can be derived using the graphical approach~\cite{Bri:71a,LinMor:82a,Varetal:88a}.
\noindent The final result  is
\begin{eqnarray}
\label{eq:crossection_jj_TI1}
\fl
\mathcal{D}^{l_c}(J_a,J_i)
=
\frac{N}{2} \;
\big( 
S_aL_a,l_i
\,
\big\vert
\big\}
\, 
S_iL_i 
\big)^2 \;
[J_a,S_i,L_i,J_i]^\half 
[l_i]^\mhalf
\nonumber
\\
\times
\vert
t(\epsilon s l_c, n_i s l_i)
\vert ^2 \;
\sum_{j_i} [j_i]
\ninej{S_a}{L_a}{J_a}{S_i}{L_i}{J_i}{s}{l_i}{j_i}^2
\,.
\end{eqnarray}
Note that the same term-independent result  can be obtained from \eref{eq:crossection_jj9}, 
thanks to the $6j$ orthogonality relations,
\begin{eqnarray}
\label{eq:jj_6j_orthogonality}
\sum_J
[J]
\sixj{j_c}{J}{J_a}{J_i}{j_i}{1}
\sixj{j_c}{J}{J_a}{J_i}{j'_i}{1}
& 
=
&
\del(j_i,j'_i)
[j_i]^{-1} 
\, ,
\end{eqnarray}
and
\begin{eqnarray}
\label{eq:sum_6j_squared}
\sum_{j_c} \; [j_c] \; \sixj{j_i}{j_c}{1}{l_c}{l_i}{s}^2 = [l_i]^{-1} 
\, ,
\end{eqnarray}
that reduce the summations over $J,j_c,j'_i$ and $j_i$ to a single sum over $j_i$.

Inserting \eref{eq:crossection_jj_TI1} in the
partial cross section formula \eref{eq:partial_cross_sect_2} and
using the one-electron reduced matrix
elements \eref{eq:red_sing_matrix}, one finds
\begin{eqnarray}
\label{eq:our_cross_section_jj}
\fl
\lefteqn{
\sigma^{TI}(J_a,J_i)
=
\frac{4\pi^2\omega}{3c}
 \sum_{l_c} [l_c] \threej{l_c}{1}{l_i}{0}{0}{0}^2
\big(\, \epsilon l_c 
\,\vert\, r \,\vert\,
n_i l_i \,\big)^2 \;
[J_a, S_i, L_i] \; }
\\ & \times &
N \;
\big(\,S_aL_a,l_i \,\big\vert\big\}\,S_iL_i\,\big) ^2 \; 
\sum_{j_i}
[\,j_i\,]
\ninej{S_a}{L_a}{J_a}{s}{l_i}{j_i}{S_i}{L_i}{J_i}^2 
. \nonumber
\end{eqnarray}
Knowing the relation (\ref{eq:from9jto6j_spec}), one realizes that the term-independent cross section of 
Pan and Starace \eref{eq:PS_eq13} is fully recovered. 
However
this expression emerges naturally from the $(jj)J$-coupling analysis, without 
calling for the knowledge of the angular momentum relation \eref{eq:from9jto6j_gen}. 

\section{Conclusion}

We have shown that the ``surprising'' agreement raised recently by Blondel {\it et al} ~\cite{Bloetal:06a,Bloetal:01a}
between the ``standard'' and the Cox-Engelking-Lineberger formulae when estimating the fine structure photodetachment 
relative intensities is understood from the important work of Pan and Starace~\cite{PanSta:93a}.
The bridge between the two formalisms can be resumed through a rather simple and useful angular momentum
relation that, to the knowledge of the authors, has never been published as such in its explicit form.
More important, the present work, adopting the irreducible tensorial expression of second quantization operators, reproduces 
Pan and Starace's parametrization of the photodetachment cross section. It provides 
an elegant and natural way to link Pan and Starace's approach (including the ``standard'' formula) 
with the fractional parentage Cox-Engelking-Lineberger formula in the term-independent approximation. It unifies the two 
formalisms through a ``simple'' recoupling of the spherical tensorial second quantized form of the E1 transition operator, 
from $(SL)J$ to $(jj)J$.

\ack

The Fonds de la Recherche Scientifique de Belgique FRFC Convention)
and the Communaut\'e fran\c{c}aise de Belgique (Actions de Recherche
Concert\'ees) are acknowledged for their financial support.

\vspace*{1cm}

\newpage
%
%
\appendix
\section{}
\label{app1} 
Analytically, the following equation holds:
\begin{eqnarray}
\label{eq:9j6j}
\displaystyle
\sum_X [X]
\ninej{a}{f}{X}{d}{q}{e}{p}{c}{b}^2
=
\sum_{Y} [Y]
\sixj{a}{b}{Y}{c}{d}{p}^2
\sixj{c}{d}{Y}{e}{f}{q}^2
\,
\end{eqnarray}
and is hereafter demonstrated graphically. The squared $9j$-symbols are joined to a $12j$
symbol by removing the sum, the momentum $X$ and the factor $[X]$, and connecting the loose ends, 
\\
\vspace*{0.5cm} 
$ \displaystyle \sum_X [X] \ninej{a}{f}{X}{d}{q}{e}{p}{c}{b}^2
=
\displaystyle
\sum_X
[X]
$
\begin{minipage}[c]{.33\textwidth}
\includegraphics{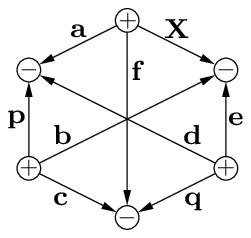}
\includegraphics{figure1.eps}
\end{minipage}
$
=
$
\begin{minipage}[c]{.25\textwidth}
\includegraphics{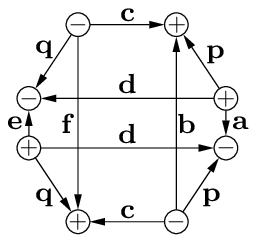}
\,.
\end{minipage} 
\\[\baselineskip]
The two pairs of momenta $(c,d)$ are cut and rejoined by a
new momentum $Y$ to obtain:
\vspace*{1cm}
\\
$
=
\displaystyle
\sum_{Y} [Y]
$
\begin{minipage}[c]{.38\textwidth}
\includegraphics{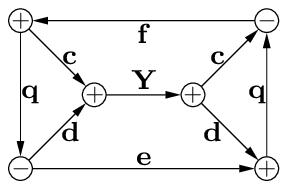}
\includegraphics{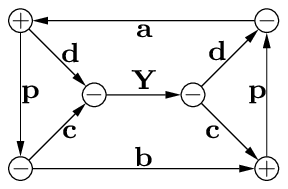}
\,.
\end{minipage}
\\[\baselineskip]
These two diagrams are cut in the middle through three momenta to get four $6j$-symbols, 
\vspace*{0.5cm}
\\ $ = \displaystyle \sum_{Y} [Y] $ \begin{minipage}[c]{.70\textwidth}
\includegraphics{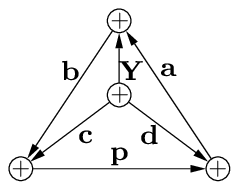}
\includegraphics{figure5.eps}
\includegraphics{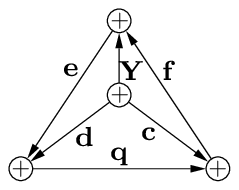}
\includegraphics{figure6.eps}
\end{minipage}
\\[\baselineskip]
$
=
\displaystyle
\sum_{Y} [Y]
\sixj{a}{b}{Y}{c}{d}{p}^2
\sixj{c}{d}{Y}{e}{f}{q}^2
\,.$
\vspace*{0.5cm} 
\\
The identity \ref{eq:9j6j} between the squared $9j$-symbol and the two squared 
$6j$-symbols is obtained. An equivalent expression, if applied to another set of momenta, is
\begin{eqnarray}
\label{eq:9j6j_B}
\displaystyle
\sum_j [j]
\ninej{j_4}{j_5}{j}{j_6}{j_7}{j_8}{j_1}{j_2}{j_3}^2
=
\sum_{j'} [j']
\sixj{j_4}{j_3}{j'}{j_2}{j_6}{j_1}^2
\sixj{j_2}{j_6}{j'}{j_8}{j_5}{j_7}^2
\,,
\end{eqnarray}
from which equation \eref{eq:from9jto6j_gen}  is derived using the symmetry properties of $6j$ and $9j$ symbols.

Note that \ref{eq:9j6j} is a special case of equation~(33)/sect.12.2 of Varshalovich {\it al.}~\cite{Varetal:88a}
\begin{eqnarray}
\label{eq:Var_eq33}
\displaystyle
\fl
\sum_X [X]
\ninej{a}{f}{X}{d}{q}{e}{p}{c}{b} 
\ninej{a}{f}{X}{h}{r}{e}{s}{g}{b} 
\\
=
\sum_{Y} [Y]
\sixj{a}{b}{Y}{c}{d}{p} \sixj{c}{d}{Y}{e}{f}{q}
\sixj{e}{f}{Y}{g}{h}{r} \sixj{g}{h}{X}{a}{b}{s}
\nonumber \\
= (-1)^{-p + q - r + s} 
\left\{
\begin{array}{cccc}
-   &  a   &   d   &   p \\
f   &  -     &  q &   c  \\
g   &  s   &  -    &    b   \\
r &  h   &  e &   -    
\end{array} \right\} 
\,, \nonumber
\end{eqnarray}
that becomes, for $ h=d, s=p, r=q, g=c$ 
\begin{eqnarray}
\label{eq:Var_eq33_sp}
\displaystyle
\fl
\sum_X [X]
\ninej{a}{f}{X}{d}{q}{e}{p}{c}{b}^2
=
\sum_{Y} [Y]
\sixj{a}{b}{Y}{c}{d}{p}^2 \sixj{c}{d}{Y}{e}{f}{q}^2
\nonumber \\
= 
\left\{
\begin{array}{cccc}
-   &  a   &   d   &   p \\
f   &  -     &  q &   c  \\
c   &  p   &  -    &    b   \\
q &  d   &  e &   -    
\end{array} \right\} 
\,. \nonumber
\end{eqnarray}
\newpage
%
%

\end{document}